\documentclass[pop,preprint,aip,amsmath,amssymb]{revtex4-1}

\usepackage[margin=3.0cm]{geometry}
\usepackage{subfigure}
\usepackage{graphicx}
\usepackage{ulem}
\usepackage{amssymb}
\usepackage[utf8x]{inputenc}
\usepackage{amsmath}
\usepackage{listings}

\begin{document}

\title {Boundary conditions for the solution of the 
3-dimensional Poisson equation in open metallic enclosures\footnote{Copyright (2015) American Institute of Physics. This article may be downloaded for personal use only. Any other use requires prior permission of the author and the American Institute of Physics. The article was published at Phys. Plasmas 22, 093119 (2015).}}

\vspace{1.2in}
\author{Debabrata Biswas, Gaurav Singh and Raghwendra Kumar}

\affiliation{Bhabha Atomic Research Centre, Mumbai 400085}

\begin{abstract}
Numerical solution of the Poisson equation in metallic enclosures, open
at one or more ends, is important in many practical situations such as 
High Power Microwave (HPM) or photo-cathode devices. It 
requires imposition of a suitable boundary condition at the open end. In this paper, 
methods for solving the Poisson equation are investigated for 
various charge densities and aspect ratios of the open ends. It is found that 
a mixture of second order and third order local asymptotic 
boundary condition (ABC) is best suited for large aspect ratios
while a proposed  non-local matching method, based on the solution of 
the Laplace equation, scores well when the aspect ratio 
is near unity for all charge density variations, including 
ones where the centre of charge is close to an open end or the charge
density is non-localized. The two methods complement each other and can be used
in electrostatic calculations where the computational 
domain needs to be terminated at the open boundaries of the 
metallic enclosure. 
\end{abstract}


\date{\today}

\maketitle
\newcommand{\be}{\begin{equation}}
\newcommand{\ee}{\end{equation}}
\newcommand{\bea}{\begin{eqnarray}}
\newcommand{\eea}{\end{eqnarray}}
\newcommand{\Tbar}{{\bar{T}}}
\newcommand{\ep}{{\cal E}}
\newcommand{\Lop}{{\cal L}}
\newcommand{\DB}[1]{\marginpar{\footnotesize DB: #1}}
\newcommand{\q}{\vec{q}}
\newcommand{\kt}{\tilde{k}}
\newcommand{\Lopn}{\tilde{\Lop}}
\newcommand{\noi}{\noindent}
\newcommand{\ovn}{\bar{n}}
\newcommand{\ovx}{\bar{x}}
\newcommand{\ovE}{\bar{E}}
\newcommand{\ovV}{\bar{V}}
\newcommand{\ovU}{\bar{U}}
\newcommand{\ovJ}{\bar{J}}
\newcommand{\calE}{{\cal E}}
\newcommand{\ovphi}{\bar{\phi}}
\newcommand{\oveps}{\bar{\ep}}

\newcommand{\PR}[1]{{Phys.\ Rep.}\/ {\bf #1}}
\newcommand{\PRL}[1]{{Phys.\ Rev.\ Lett.}\/ {\bf #1}}
\newcommand{\PP}[1]{{Phys.\ Plasmas\ }\/ {\bf #1}}
\newcommand{\JAP}[1]{{J.\ Appl.\ Phys.}\/ {\bf #1}}
\newcommand{\JCP}[1]{{J.\ Comput.\ Phys.}\/ {\bf #1}}
\newcommand{\CPC}[1]{{Comput.\ Phys.\ Commun.}\/ {\bf #1}}

\section{Introduction}
\label{sec:intro}
Open computational boundaries pose a challenge for both time-dependent
and time-independent problems in fields as diverse as electromagnetics,
quantum mechanics, fluid dynamics and biological systems. 
In electromagnetics, examples of devices with one or more open 
ends include the Virtual Cathode Oscillator or the Klystron \cite{granat,cairns}, 
those involving photo-cathodes or a charge-particle beam in a conducting pipe.
Their simulation using a Particle-in-Cell (PIC) code requires the imposition of 
artificial boundary conditions at the open ends of the computational 
domain. For time-varying electromagnetic fields, the Perfectly Matched Layer (PML) technique 
is commonly adopted and provides a viable non-reflecting termination of
the computational domain at an extra cost \cite{berenger,taflove}. 
For electrostatic fields, the Poisson equation

\be
\nabla^2 V({\bf r}) = -\rho({\bf r})/\epsilon_0 \label{poisson}
\ee

\noindent
needs to be solved with specified boundary conditions 
and for open ends, special techniques need to be adopted \cite{konrad}. 
Here $V$ is the electrostatic potential, $\rho$ is the charge density 
and $\epsilon_0$ is free space permittivity.

Depending on the physical situation being modeled, various 
scenarios may arise.
When the problem of interest comprises of a charge distribution
that is sufficiently isolated from other objects, 
the boundary condition at infinity 
can be implemented by choosing the computational
domain to be spherical and applying a suitable artificial
boundary condition at the surface.  Alternately, the free-space 
Green's function
can be used to evaluate the field inside the computational domain.
In such situations, efficient methods exist that limit the
computational cost \cite{isf}.

Quite often, in addition to charges, the computational domain may consist of 
metallic objects where additional boundary conditions need to be imposed.
When the electrostatic field outside the metallic objects is of interest,
the computational boundary can be a chosen to be 
sphere or a cube with a suitable artificial boundary condition. 
The local Asymptotic Boundary Conditions (ABC) \cite{bayliss,mittra},
the Dirichlet to Neumann (DtN) map \cite{han,keller} and hybrid methods such as
the boundary relaxation/potential method \cite{silvester,Miller,zwang} 
and the boundary integral method \cite{bim} are particularly 
useful in such situations. 

\begin{figure}[htb!]
     \begin{center}
            \includegraphics[width=0.35\textwidth, angle=0]{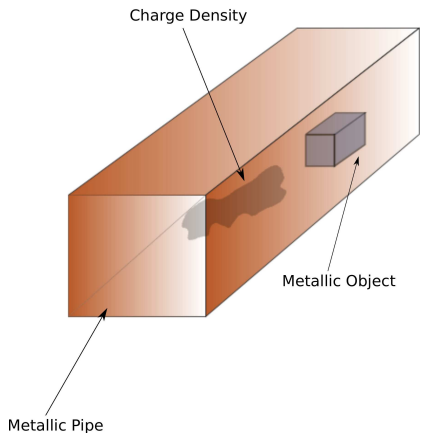}
\end{center}

\caption{A schematic of an open pipe having charges and a metallic structure inside.}
\label{fig:pipe}
\end{figure}

Finally, the region of interest may lie {\it inside}  open 
metallic objects such as a guide tube or an open pipe 
(see Fig~\ref{fig:pipe}) enclosing
charges and perhaps other metallic structures such as a
cathode or anode.
In such cases, in order to minimize computational resources, 
it is natural to limit the computational domain to 
the extent of the outer open metallic enclosure  
and impose a suitable approximate boundary
condition at the open surfaces (such as the two ends of a pipe or
guide tube). We shall limit ourselves to this last category 
of problems in this paper.

To the best of our knowledge, there are few boundary conditions 
available in open literature that can be directly applied in
a finite difference scheme when the computational
domain is truncated at one or more open ends. 
The simplest of these is the first order
Asymptotic Boundary Condition (ABC1), 

\be
\frac{\partial V}{\partial r} + \frac{V}{r} = 0 \label{eq:ABC1}
\ee

\noindent
which can be easily implemented at an open boundary $x_i =$ constant, 
thereby allowing a solution on a Cartesian grid using the finite
difference technique. ABC1
is based on the general solution \cite{jackson}

\be
V(r,\theta,\phi) = \sum_{l=0}^{\infty} \sum_{m=-l}^{m=l} \frac{B_{lm}}{r^{l+1}} Y_{l,m}(\theta,\phi)
  \label{eq:Lapsol}
\ee

\noi
in the charge-free region outside the computational domain.
Here $Y_{l,m}(\theta,\phi)$ are the spherical harmonics and $B_{lm}$ are
unknown coefficients. Equation (\ref{eq:ABC1}) follows on noting that

\be
\frac{\partial V}{\partial r} + \frac{V}{r} = \mathcal{O}(1/r^3)
\ee

\noi
and, as a first approximation, the right hand side can be set to zero. 
Successive boundary
conditions can be similarly derived \cite{bayliss,mittra}. 
For example, the second order
asymptotic boundary condition (ABC2) is

\be
\frac{\partial^2 V}{\partial r^2} + \frac{4}{r} \frac{\partial V}{\partial r}
+ \frac{2 V}{r^2}  = \mathcal{O}(1/r^5) \simeq 0.
\ee

\noi
In general, there exists a hierarchy of such boundary conditions
which can be expressed as 

\be
\prod_{j=1}^{n} ({\partial  \over \partial r} + {(2j - 1) \over r}) V = 0
\label{eq:ABC_general}
\ee

\noi
which represent the $n^{th}$ order asymptotic boundary condition, ABCn.
A local implementation however requires the use of 
lower order ABC, thereby diluting the accuracy. 
For example, a local implementation of the
second order method (ABC2) requires the use of ABC1 
in order to compute mixed derivatives $\partial^2 V/\partial x_i\partial x_j$
at an open boundary $x_i =$ constant.
We shall show that ABC2 generally delivers 
good results unless the charge density variation perpendicular to the
open boundary is low, in 
which case the method is found to be inappropriate.
For large aspect ratio open boundaries however, 
the Asymptotic Boundary Condition remains stable and
consistent and as we shall show, a mixture of ABC2 and ABC3
can help reduce errors.

Apart from the local asymptotic boundary conditions,
non-local hybrid methods can be applied to an open
pipe geometry even though they are resource intensive. 
The  boundary relaxation/potential technique for solving  
the Poisson equation relies on iteratively correcting a solution of the 
Poisson equation with an assumed potential at the open boundary.

Both the asymptotic boundary condition (ABC) and the 
boundary potential method can be directly implemented in an
open pipe or guide-tube geometry without expanding the computational
domain. We propose here an alternate non-local method (referred to as Method-1
hereafter) that complements ABC when the aspect ratio of the 
open surface is near unity. It scores very well over ABC when the centre of 
charge is near an open end or when the charge distribution is non-localized.
It relies on matching the potential to the
solution of the Laplace equation at the open boundary using
$(l_{max} + 1)^2$ points where the sum over $l$ in Eq.~(\ref{eq:Lapsol})
is restricted to $l_{max}$. In effect it uses less than $0.5\%$ points
on the open surface and gives consistent results that are generally
better than the methods discussed above.
It has been studied using Finite Element Method (FEM)
\cite{chari1,chari2} but its application using Finite Difference
is limited to a 2-dimensional situation \cite{hammel} perhaps on account of
convergence issues. 

In section \ref{sec:boundary}, we outline the proposed
Method-1 that we shall adopt for aspect ratios close to 
unity. Thereafter, we shall review the 
implementation of the second and third order
Asymptotic Boundary Conditions (ABC2 and ABC3) in section \ref{sec:ABC}
and outline the boundary potential method in section \ref{sec:bpm}
for the sake of comparison.
Section \ref{sec:numerics} deals with the numerical
results for charge densities in an open 
rectangular pipe, a problem for which, the exact solution 
is known.

\section{Non-local Boundary truncation using Laplace solution (Method-1)}
\label{sec:boundary}

We propose here a method that is especially useful when the 
aspect ratio of the open end is near unity and charges are 
near the open boundary.

Consider an open boundary $x = x_N$ in a 3-dimensional Cartesian grid
where $x_1,x_2,\ldots,x_N$ are equispaced points along the $X-$axis.
We need to specify the boundary potential $V_{N,j,k}$ at the open 
boundary $x = x_N$ in order to solve the Poisson equation inside
the computational domain. In this section,
we determine a boundary truncation scheme whereby $\{V_{N,j,k}\}$  
can be expressed in terms of $\{V_{N-1,j,k}\}$. A self-consistent
iteration scheme can then be used to find the potential in the
region of interest.

Using central difference, the normal derivative of the potential at 
the open boundary is

\be
{\frac{\partial V}{\partial x}}|_{x=x_N}  = \frac{(V_{N+1,j,k} - V_{N-1,j,k})}{2h_x} \label{eq:FD}
\ee

\noindent
where $h_x$ is the spacing between points along the $X$-direction
(similarly, $h_y$ and $h_z$ denote spacing along $Y$ and $Z$ directions).
Using the Laplace solution (Eq.~\ref{eq:Lapsol}) 
to evaluate $(\partial V/\partial x)_{|_{x=x_N}} $, and
$V_{N+1,j,k}$, a system of linear equations can be set up to determine the unknown
coefficients $B_{lm}$ in terms of $V_{N-1,j,k}$. To this end, note that

\bea
\frac{\partial V}{\partial x} & = &\sum_{l=0}^{\infty} \sum_{m=-l}^l B_{lm} \left[-\frac{(l+1)x}{r^{l+3}} 
Y_{l,m}(\theta,\phi)~ \right. \nonumber \\ 
& + & \left.
 \frac{1}{r^{l+1}} \left\{ \frac{\partial Y_{l,m}}{\partial\theta} \frac{\partial\theta}{\partial x} + \frac{\partial Y_{l,m}}{\partial \phi}\frac{\partial \phi}{\partial x}\right\}\right]
\label{eq:Deriv}
\eea

\noi
where 

\bea
\frac{\partial \theta}{\partial x} & = & \frac{zx}{r^2\sqrt{x^2 + y^2}} \\
\frac{\partial \phi}{\partial x} & = & -\frac{y}{x^2 + y^2} 
\eea

\noi
and

\bea
\frac{\partial Y_{l,m}}{\partial \phi}&=&\iota m Y_{l,m}(\theta,\phi) \\
\frac{\partial Y_{l,m}}{\partial \theta}&=&  \sqrt{(l-m)(l+m+1)}~ 
e^{-\iota\phi} Y_{l,m+1}(\theta,\phi)    \nonumber \\
& + & m\cot\theta~ Y_{l,m}(\theta,\phi).
\eea 

\noi
Here, ($r,\theta,\phi)$ are the spherical polar co-ordinates.
Using the above, the system of equations can thus be expressed as

\be
\begin{split}
\sum_{l=0}^{\infty} \sum_{m=-l}^l & B_{lm}  \left[ \frac{Y_{l,m}(\theta_{N+1},\phi_{N+1})}{r_{N+1}^{(l+1)}} - 2h_x \left\{-\frac{(l+1)}{r_N^{l+3}} x_N Y_{l,m}(\theta_N,\phi_N)~+~ \right. \right. \\
& \left. \left. \frac{1}{r_N^{l+1}} \left( \frac{\partial Y_{l,m}}{\partial\theta} 
\frac{\partial\theta_N}{\partial x}
  +  \frac{\partial Y_{l,m}}{\partial \phi}
\frac{\partial \phi_N}{\partial x}  \right)_{|_{(r_N,\theta_{N},\phi_{N})}} \right\} \right] 
 = V_{N-1,j,k} \label{eq:linset}
\end{split}
\ee
\noi
In practice, the sum over $l$ must be truncated at $l=l_{max}$ and the number of points $\{j,k\}$ on the open
boundary $x = x_N$ chosen to equal the number of unknown coefficients $B_{lm}$. It is easy to verify that
truncation at $l = l_{max}$ leads to $N_{max} = (l_{max} + 1)^2$  number of unknowns. Thus  $N_{max}$
points must be chosen appropriately on the open boundary.

The system of equations in Eq.~(\ref{eq:linset}) can be solved to yield $B_{lm}$ which in 
turn can be used to find the potential $\{V_{N,j,k}\}$ in terms of $\{V_{N-1,j,k}\}$. Similarly,
for a boundary on the left ($x = x_1$), $\{V_{1,j,k}\}$ can be expressed in terms
of $\{V_{2,j,k}\}$. Thus, the potential at all points between $x_1$ and $x_N$ can be 
updated using a standard Poisson solver.

The numerical results using Method-1 are presented in
section \ref{sec:numerics}.
In the finite difference implementation however, there are convergence 
issues depending on origin. Nevertheless, the domain of convergence can be 
determined easily when the aspect ratio of the open face is
between $1/4$ and $4$. 

\begin{figure}[htb!]
     \begin{center}
            \includegraphics[width=0.35\textwidth, angle=0]{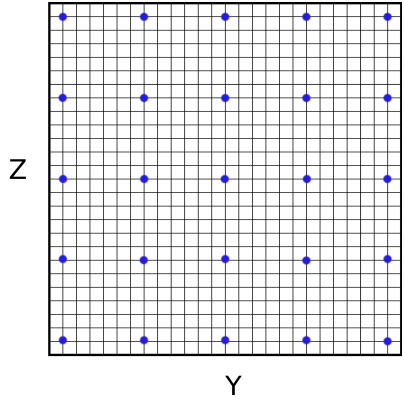}
\end{center}

\caption{A $27 \times 27$ grid on open surface $X=$constant along with 
the 25 points (solid dots) chosen for matching the solution. In the rest of the 
paper, a $41 \times 41$ grid is used.}  
\label{fig:choose}
\end{figure}

Importantly, for most problems, $l_{max} = 4$ is sufficient for
implementing the open boundary. Thus a total of 25 points on an open surface
need to be chosen out of $\mathcal N^2$ points 
where $\mathcal N$ is the average number
of points along the edge of an open face needed by the Poisson solver. 
We choose an implementation shown in Fig~(\ref{fig:choose}).

\section{Implementation of the Asymptotic Boundary Conditions (ABC)}
\label{sec:ABC}

We shall briefly outline the implementation of the well known
Asymptotic Boundary Condition in this section. We shall assume
the open face to be $x = $ constant without any loss of
generality.

\subsection{ABC1}
\label{subsec:ABC1}

The first order asymptotic boundary condition 

\be
({\partial \over \partial r} + {1\over r})V = 0  
\ee
\noi
can be expressed as 

\be
V_x {\partial x \over \partial r} + V_y {\partial y \over \partial r} +
V_z {\partial z \over \partial r} + {V \over r} = 0.
\ee

\noindent
Using $\partial x / \partial r = x/r$,  $\partial y / \partial r = y/r$
and  $\partial z / \partial r = z/r$, the above equation can be
expressed as

\be
V_x = -{V\over x} - {y \over x}V_y  - {z \over x}V_z
\ee

\noi
which can be discretized at $x = x_{N-1}$ to yield

\be
V_{N} = V_{N-2} - (2h_x) \left[{V\over x} + {y \over x}V_y  + {z \over x}V_z\right]_{N-1} 
\ee

\noi
where $ x_N - x_{N-1} = x_{N-1} - x_{N-2} = h_x$. Thus the potential at the 
face $x = x_N$ can be expressed in terms of the potential and its
derivative inside the computational domain.

\subsection{ABC2}
\label{subsec:ABC2}

The second order asymptotic boundary condition

\be
{\partial^2 V \over \partial r^2} + {4 \over r}{\partial V \over \partial r}
 + { 2V \over r^2} = 0
\ee

\noi
can be similarly implemented at an open surface at $x = x_N$ on
expressing ${\partial^2 V / \partial r^2}$ as

\be
V_{rr} = V_{xx} {x^2 \over r^2} + V_{yy} {y^2 \over r^2} + V_{zz} {z^2 \over r^2} 
+ V_{xy} {2xy \over r^2} + V_{xz} {2xz \over r^2} + V_{yz} {2yz \over r^2}.
\ee

\noi
Thus,

\be
V_x + {x \over 4}V_{xx} = -{V \over 2x}  + {y^2 \over 4x} V_{yy} + {z^2 \over 4x} V_{zz}
 + {yz \over 2x} V_{yz}
\ee

\noi
so that on discretizing at $x = x_{N-1}$, we have

\be
V_{N} = \frac{1}{1+\frac{x}{2h_x}}\left[V_{N-2} + \frac{x}{2h_x}\left(2V_{N-1} - V_{N-2}\right)\right]  
+ 2h_x \left[ -{V \over 2x} + {y^2 \over 4x} V_{yy} + 
{z^2 \over 4x} V_{zz} + {yz \over 2x} V_{yz} \right ]_{N-1}
\ee

\noindent
where $[\ldots]_{N-1}$ denotes discretization at $x = x_{N-1}$.

\subsection{ABC3}
\label{subsec:ABC3}

Implementation of the third order asymptotic boundary condition

\be
{\partial^3 V \over \partial r^3} + {9 \over r} {\partial^2 V \over \partial r^2}
+ {18 \over r^2} {\partial V \over \partial r} + {6V \over r^3} = 0
\ee

\noi
at the $x = x_N$ face requires $V_{rrr}$ to be expressed in terms of 
the partial derivatives in cartesian co-ordinates:

\be
\begin{split}
V_{rrr} = & V_{xxx} {x^3 \over r^3} + V_{xxy} {3x^2 y \over r^3} + V_{xxz} {3x^2 z \over r^3} 
+ V_{yyy} {y^3 \over r^3} + V_{yyx} {3xy^2 \over r^3} + V_{yyz} {3y^2 z \over r^3} \\
& + V_{zzz} {z^3 \over r^3} + V_{zzx} {3xz^2 \over r^3} + V_{zzy} {3yz^2 \over r^3} 
+ V_{xyz} {6 xyz \over r^3}. 
\end{split}
\ee

\noi
Together with the expressions for $V_{rr}$ and $V_r$,
$V_x$ can be similarly obtained such that $V_N$ is expressed
in terms of the potential and its derivatives at interior points.
The analysis above can be similarly generalized 
for faces $y,z$=constant.

\section{Truncation using the boundary potential method (BPM)}
\label{sec:bpm}

Apart from the proposed Method-1 and the Asymptotic Boundary
Conditions, the Boundary Potential Method can also be directly
applied when the computational domain is truncated at the
open face. We shall review the implementation briefly and use
it in the next section for the sake of comparison.

 Consider a metallic rectangular wave-guide with two open faces.
It may contain some metallic 
structure (see Fig.~(\ref{fig:pipe})) or a charge distribution $\rho({\bf r})$ or both. 
We need to solve Poisson equation (Eq.~(\ref{poisson})
with boundary conditions $V=V_D|_S$, where $V_D$ is the specified 
potential on the metallic surfaces and $V({\bf r})= Q/(4 \pi \epsilon_0 {\bf r})$
as ${\bf r}\rightarrow \infty$. Here $Q$ is the sum of all charges inside the domain 
(in this case, the open wave-guide) consisting of charge density $\rho$ and also the 
surface charges present on all surfaces.

To solve this problem, the following steps need to carried out:

\begin{enumerate}
\item
Poisson equation in the domain of interest is solved with an assumed potential (e.g. $V = 0$) 
at the open boundaries and the specified potential $V_D$ on the remaining surfaces
using a standard numerical procedure. The solution $V_0({\bf r})$ obtained with the assumed boundary potential is
clearly different from the desired solution and gives rise to surface charges at the open
boundary. The solution can be corrected iteratively as described in following steps.

\item
The screening surface charge 
density $\sigma_{os}$ at the open surfaces is calculated by taking the 
normal derivative of the  potential $V_0({\bf r})$

\be
\sigma_{os}= -\epsilon_0 {\bf n}\cdot \nabla V_0({\bf r}).
\ee

\item
The boundary potential $V^{k}_{os}$ ($k=0$) at open surface due to all the screening 
charges is calculated using the free space Green's function: 

\be
V^{0}_{os}({\bf r})=-{\frac{1}{\epsilon_0}} \int {\int{ dS' G({\bf r}|{\bf r'}) \sigma_{os}({\bf r'})}}
\ee

\item
\label{psi}
Next, Laplace equation $\nabla^2 \psi=0$  is solved inside the domain of interest to calculate 
the correction potential $\psi$. The boundary conditions on $\psi$ are:

\be
\psi = \begin{cases} V^k_{os} & \mbox{at open surfaces}, \\
 0 & \mbox{at all other surfaces.} \end{cases}
\ee

\item
The calculated correction potential, $\psi$, itself needs correction. 
This is so because the free space Green's function is used to calculate $V^{0}_{os}$, 
in effect ignoring the presence of all the surfaces where potential was already 
specified. For instance, in case of a metallic pipe, the presence of the wall and 
inner metallic structures (if any) is ignored.

In order to include the effect of all surfaces other than the open surfaces, the 
screening charge is calculated at these inner surfaces using the normal derivative of $\psi$:

\be
\sigma^{k}_{in}= -\epsilon_0 {\bf n}\cdot \nabla \psi.
\ee

\item

The correction in boundary potential due to these screening charges is again calculated using the free space Green's function.

\be
V^{correction}_{os}({\bf r})={\frac{1}{\epsilon_0}}\int {\int{ dS' G({\bf r}|{\bf r'}) \sigma^{k}_{in}({\bf r'})}}
\ee

\item 
\label{correct}
Corrected boundary potential is given by
\be
V^{k}_{os}=\omega V^{0}_{os}+\omega V^{correction}_{os}+ (1-\omega)V^{k-1}_{os}. 
\ee

For $0<\omega<1$, above correction formula assures convergence for any well 
resolved geometry \cite{Miller}.

\item
One needs to iterate step \ref{psi} to step \ref{correct} till $V^{k}_{os}$ converges to the
required tolerance level. The solution $V$ to  equation \ref{poisson} is given by:

\be
V({\bf r})=V_0({\bf r})+\psi({\bf r})
\ee

where $\psi({\bf r})$ is obtained as in step \ref{psi} using converged boundary 
potentials at the open surfaces while $V_0({\bf r})$ is calculated in step 1.

\end{enumerate}

The scheme discussed above is implemented using Finite Difference 
and compared with the proposed Method-1 and ABC2 in the following section.

\section{Numerical Results}
\label{sec:numerics}

In order to study the efficacy of the three boundary truncation methods
under different conditions, we shall study 
various charge densities inside a rectangular metallic pipe with
open ends having specified aspect ratios. For this problem, the exact 
solution can be easily computed \cite{pipe} using the exact Green's function.

For a rectangular  pipe of dimension $L_x$, $L_y$ and $L_z$, 
with open faces at $x = 0$ and $x = L_x$, the potential
can be expressed as \cite{pipe}

\be
\begin{split}
V&(x,y,z) =  \frac{2}{L_yL_z\epsilon_0} \sum_n \sum_m \frac{1}{\gamma_{m,n}}\sin(k_y y)\sin(k_z z) \times \\
& \iiint\limits_{-\infty}^{~~~~\infty}
e^{-\gamma_{m,n}|x-x'|} \sin(k_y y') \sin(k_z z') \rho(x',y',z')d^3r' \label{eq:exact}
\end{split}
\ee

\noi
where $k_y = m\pi/L_y$, $k_z = n\pi/L_z$, $\gamma_{m,n}^2 = \pi^2(m^2/L_y^2 + n^2/L_z^2)$.
We now define various problems based on the form of the charge density
$\rho(x,y,z)$. Note that Eq.~(\ref{eq:exact}) does not hold if there are
other metallic objects inside the pipe.

We shall test the boundary conditions essentially in two different scenarios.
In the first, we shall allow the aspect ratio of the open faces 
to be unity but allowing for variation in the length of the enclosure.
The second deals with aspect ratios beyond unity. In both cases, mesh-independence
studies have been carried out by ensuring that the average relative error (see Eq.~(\ref{eq:error}))
saturates within an acceptable limit as the size of the grid is increased.
While, the grid size at which results become mesh-independent depends on the
charge density chosen, it is generally found that a grid size of $81 \times 81 \times 81$
is adequate. All error estimates reported hereafter use this grid size, unless
otherwise mentioned.

\subsection{Unit Aspect Ratio}

We shall first consider the case where the aspect ratio $L_y/L_z = 1$.
To begin with we choose $L_x = L_y =  L_z = 1.0$m with the open
ends at $x = 0$ and $x = L_x$. The potential is computed using (a) the exact 
expression given in Eq.~(\ref{eq:exact}) with the sum truncated appropriately 
to ensure convergence
(b) the Laplace equation based non-local method of section \ref{sec:boundary} (referred
to as Method-1) with $l_{max} = 4$,
(c) the iterative Green's function based Boundary Potential Method  with $\omega = 0.5$ and
(d) the local asymptotic boundary condition ABC2. In each case a $81\times 81 \times 81$ grid
is chosen that includes the boundary points. Unless, otherwise specified, all distances
are measured in metres, the potential in volts and charge density in coulomb per cubic metre.

\begin{table}[hbt]
\vskip 0.25 in
\begin{center}
    \begin{tabular}{ | c | c| c | c| c | c| c |}
    \hline
\textbf{Density} & \multicolumn{2}{c|}{\textbf{Method-1}} &  \multicolumn{2}{c|}{\textbf{BPM}} &  \multicolumn{2}{c|}{\textbf{ABC-2}}\\
\cline{2-7}
& (Full) & (Interior) & (Full) & (Interior) & (Full) & (Interior)  \\
\hline \hline
Case-1 & ~1.26\% & 0.54\% & 17.41\% & 7.71\% & 21.31\% & 15.64\% \\
\hline
Case-2 & ~2.01\% & 0.62\% &  21.11\% & 9.06\% & 1.68\% & 1.12\%   \\
\hline
Case-3 & ~1.5\%  & 0.58\% & 18.85\% & 8.18\% & 10.58\% & 7.89\% \\
\hline
 
\end{tabular}
\caption{The average error for the three charge densities Case-1,2 and 3
for an enclosure with $L_x = L_y = L_z = 1$ having open faces at $x = 0$
and $x = L_z$
(a) the full domain excluding only the boundary points and (b) Interior points.
For the full domain, the error analysis considers $N = 79^3$ points.
Since distortion is often closest to the boundary, the error for interior domain 
is computed using $N = 67^3$ points i.e. leaving 6 additional
points along each face of the computational domain.}
\end{center}
\end{table}

\subsubsection{Uniform density along the pipe axis (Case-1)}
\label{subsec:uniform}

\begin{figure}[thb!]
     \begin{center}

        \subfigure[Exact]{
            \label{fig:1a}
            \includegraphics[width=0.3\textwidth, angle=270]{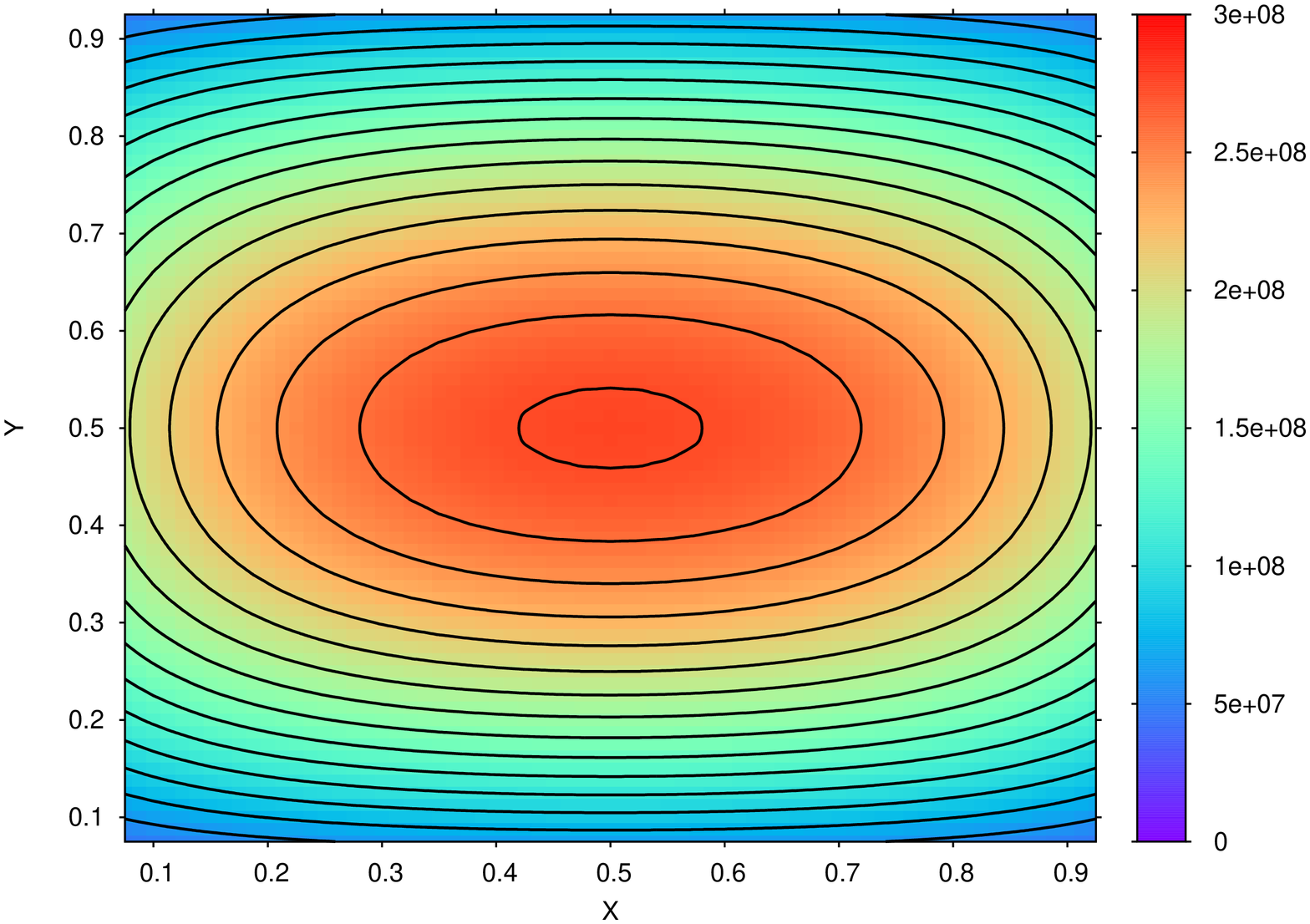}}
        \subfigure[Method-1]{
           \label{fig:1b}
           \includegraphics[width=0.3\textwidth, angle = 270]{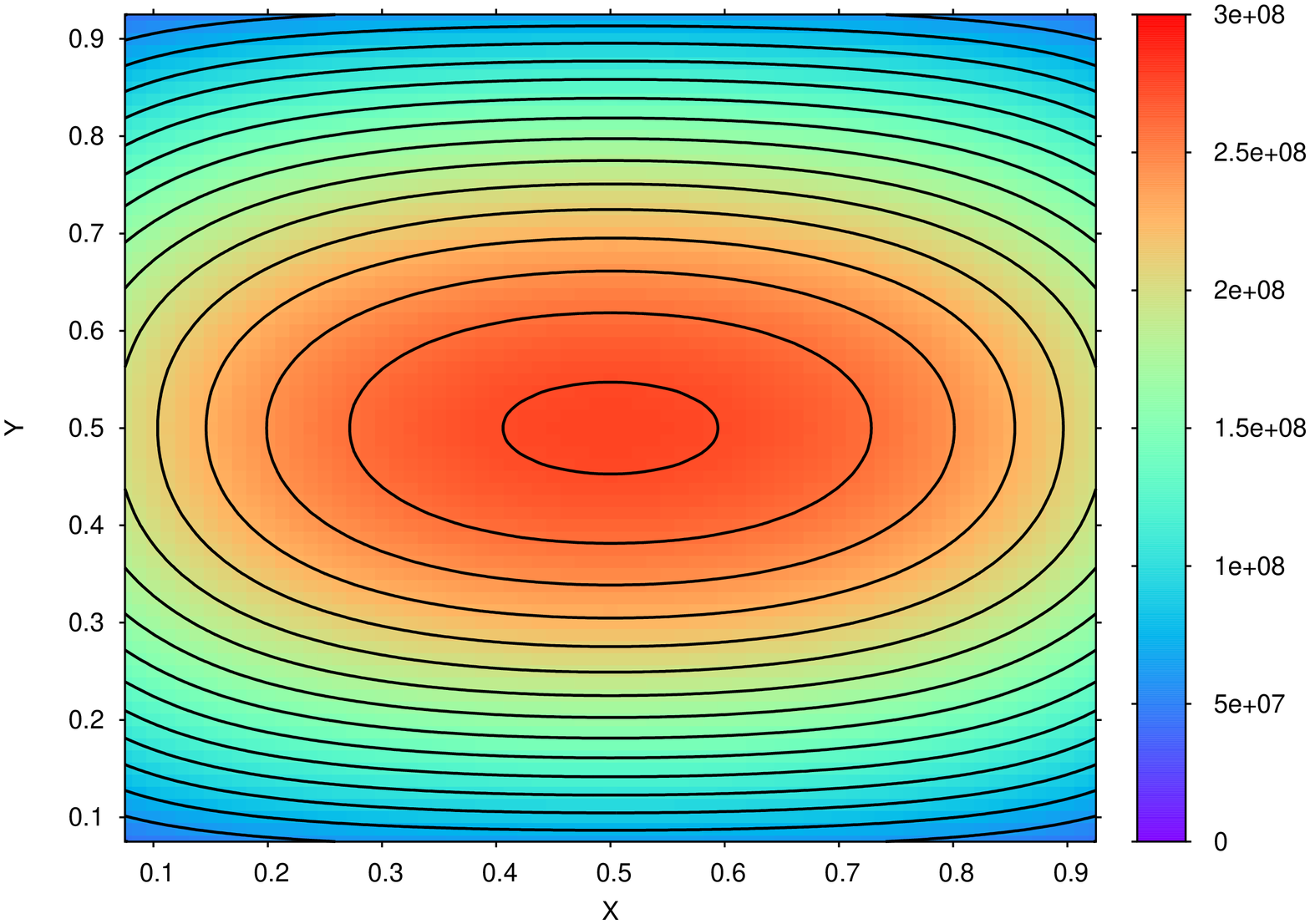}}
        \\ 
        \subfigure[BPM]{
            \label{fig:1c}
            \includegraphics[width=0.3\textwidth, angle= 270]{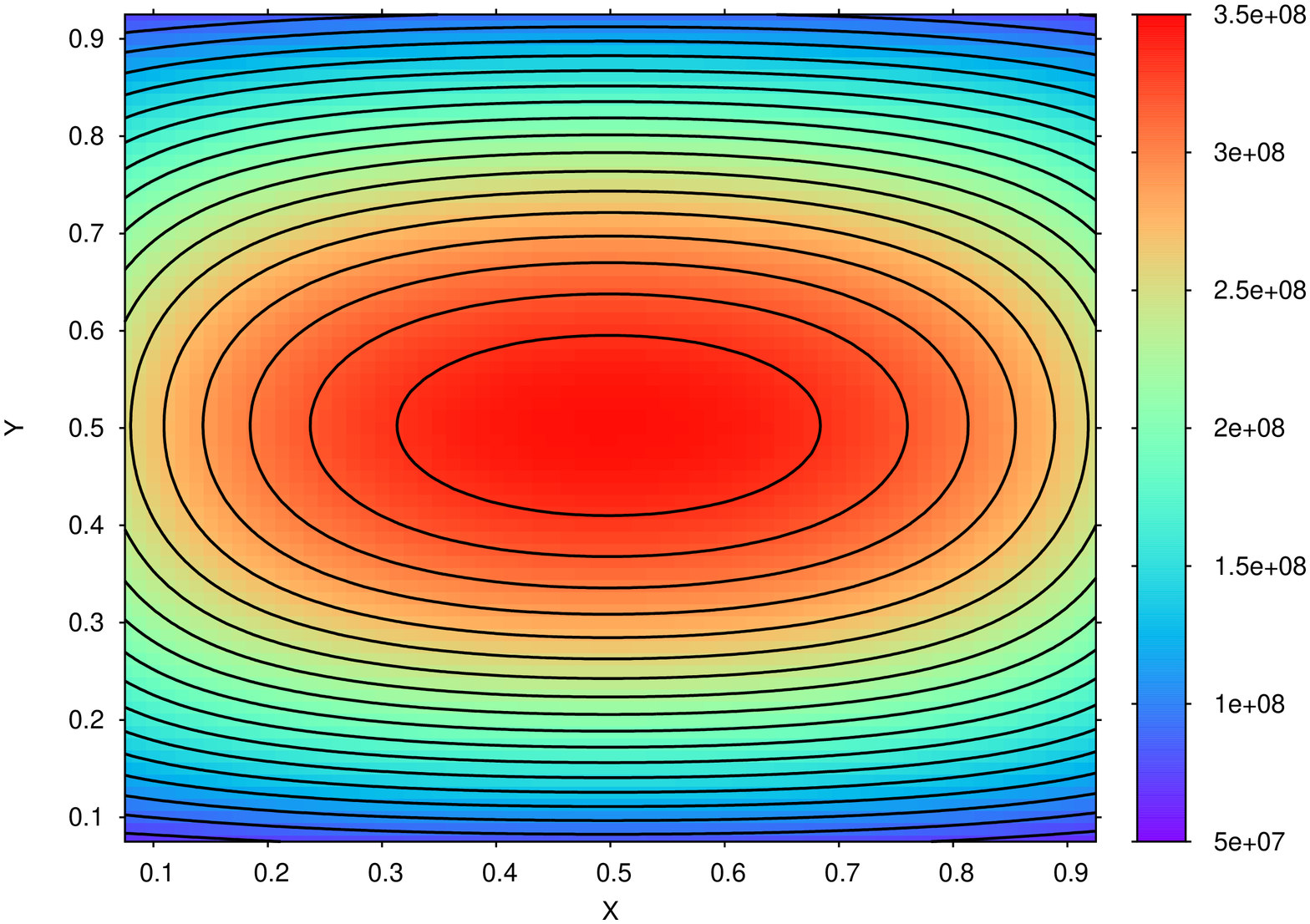}}
        \subfigure[ABC2]{
            \label{fig:1d}
            \includegraphics[width=0.3\textwidth, angle = 270]{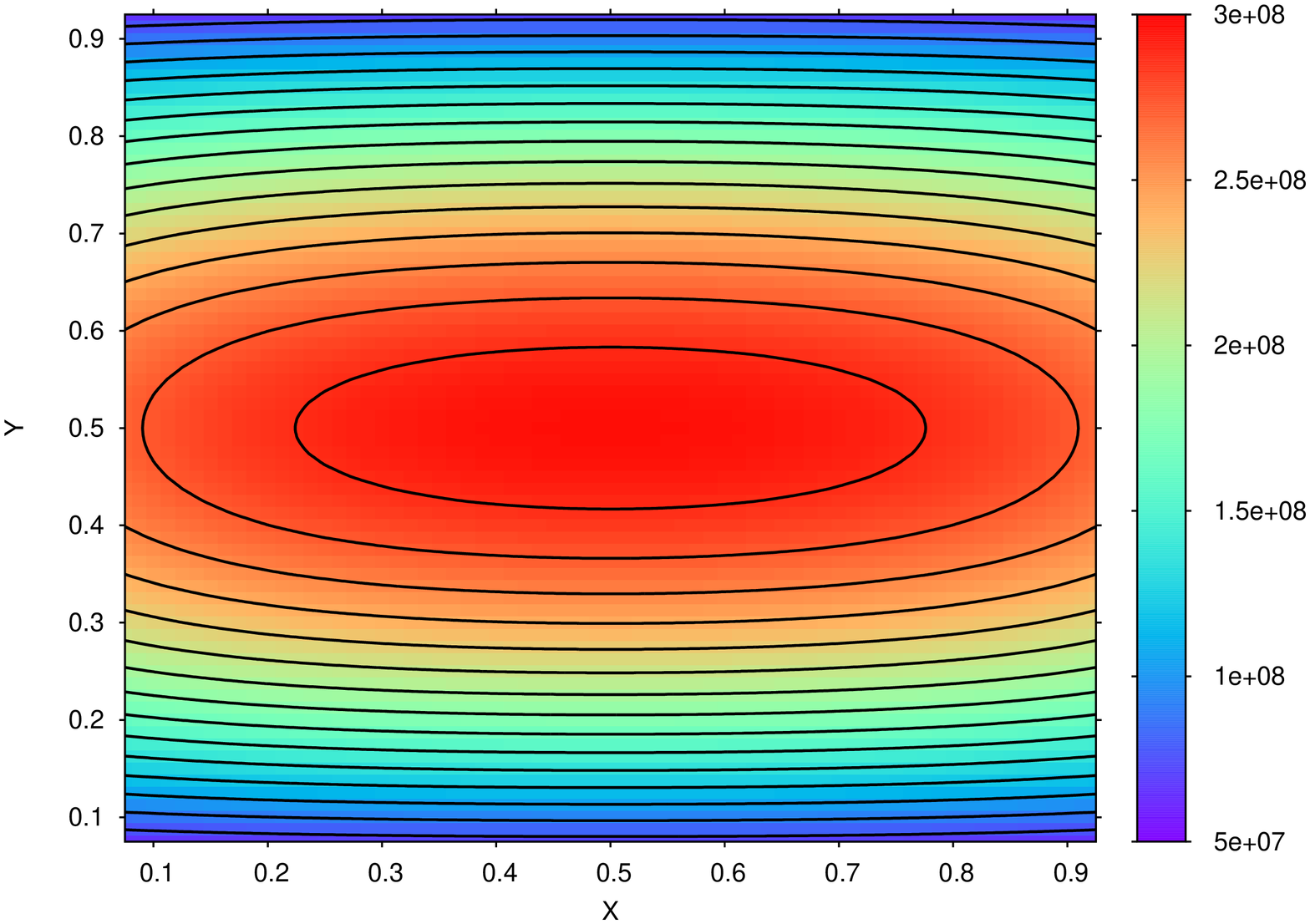}}
    \end{center}
    \caption{
        The potential $V$ for a charge density uniform in the X-direction and
parabolic in the $Y$ and $Z$ directions. A projection on the $XY$ plane $z=0.3$
is shown in each case. 
     }
   \label{fig:d1}
\end{figure}

We first choose a charge density that is uniform along the $X$-axis
but varies along the $Y$ and $Z$ directions \cite{pipe}:

\be
\rho(x,y,z) = \begin{cases} 
\left[ \frac{L_y^2}{4} - ( y - \frac{L_y}{2} )^2 \right] 
\left[ \frac{L_z^2}{4} - ( z - \frac{L_z}{2} )^2 \right]  \\
 \\
0 ~~~~~ \mbox{for}~~ x < 0~~\&~~ x > L_x . \end{cases}
\ee

\noi
The results are shown in Fig.~(\ref{fig:d1}).
Clearly, Method-1 is closest to the exact result while ABC2 performs
rather poorly in this case. A comparison of the average relative error ($\%$)

\be
\mbox{Error} = \frac{1}{N} \sum_{i,j,k}  \frac{|V_{i,j,k} - V_{i,j,k}^{exact}|}{V_{i,j,k}^{exact}} \times 100 
\label{eq:error}
\ee

\noi
is given table 1. Here $N$ is the number of points sampled and $V_{i,j,k}^{exact}$ is calculated
using Eq.~(\ref{eq:exact}). Since Method-1 and ABC results depend on the choice of
the origin, the best case relative error is provided.

\subsubsection{Two Localized Gaussian charge densities (Case-2)}
\label{subsec:2gaussian}

\begin{figure}[htb!]
     \begin{center}

        \subfigure[Exact]{
            \label{fig:2a}
            \includegraphics[width=0.3\textwidth, angle=270]{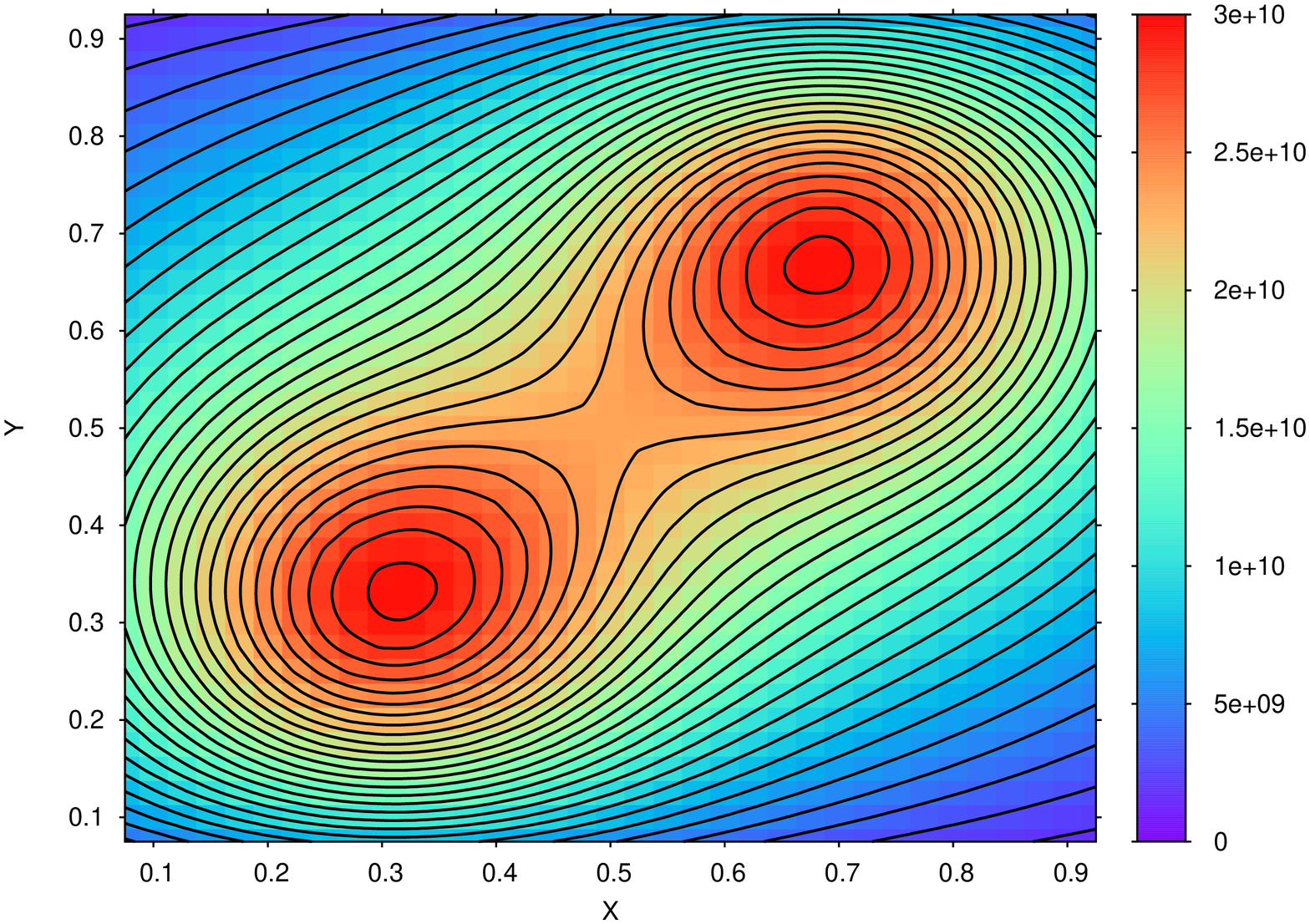}}
        \subfigure[Method-1]{
           \label{fig:2b}
           \includegraphics[width=0.3\textwidth, angle = 270]{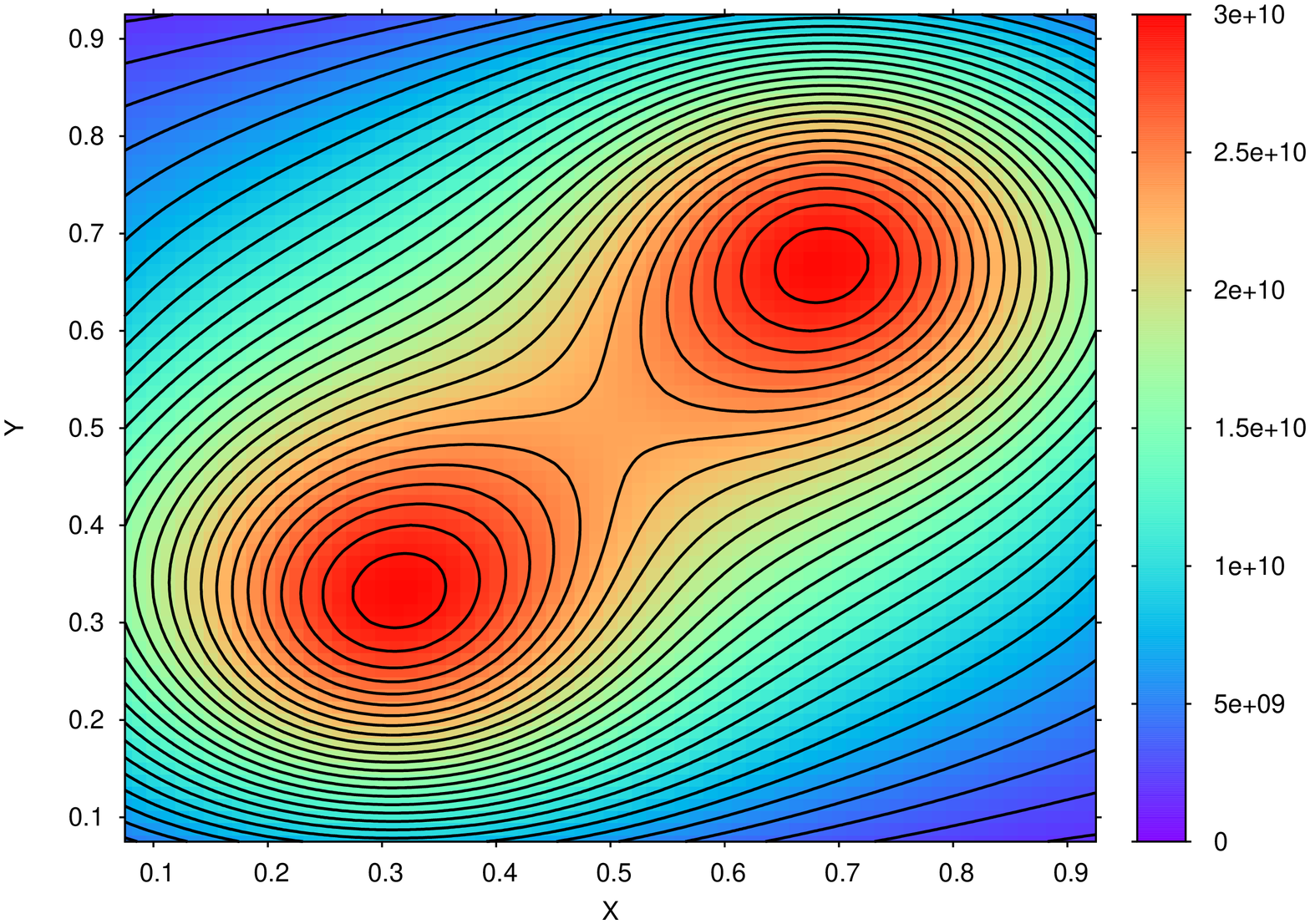}}
        \\ 
        \subfigure[BPM]{
            \label{fig:2c}
            \includegraphics[width=0.3\textwidth, angle= 270]{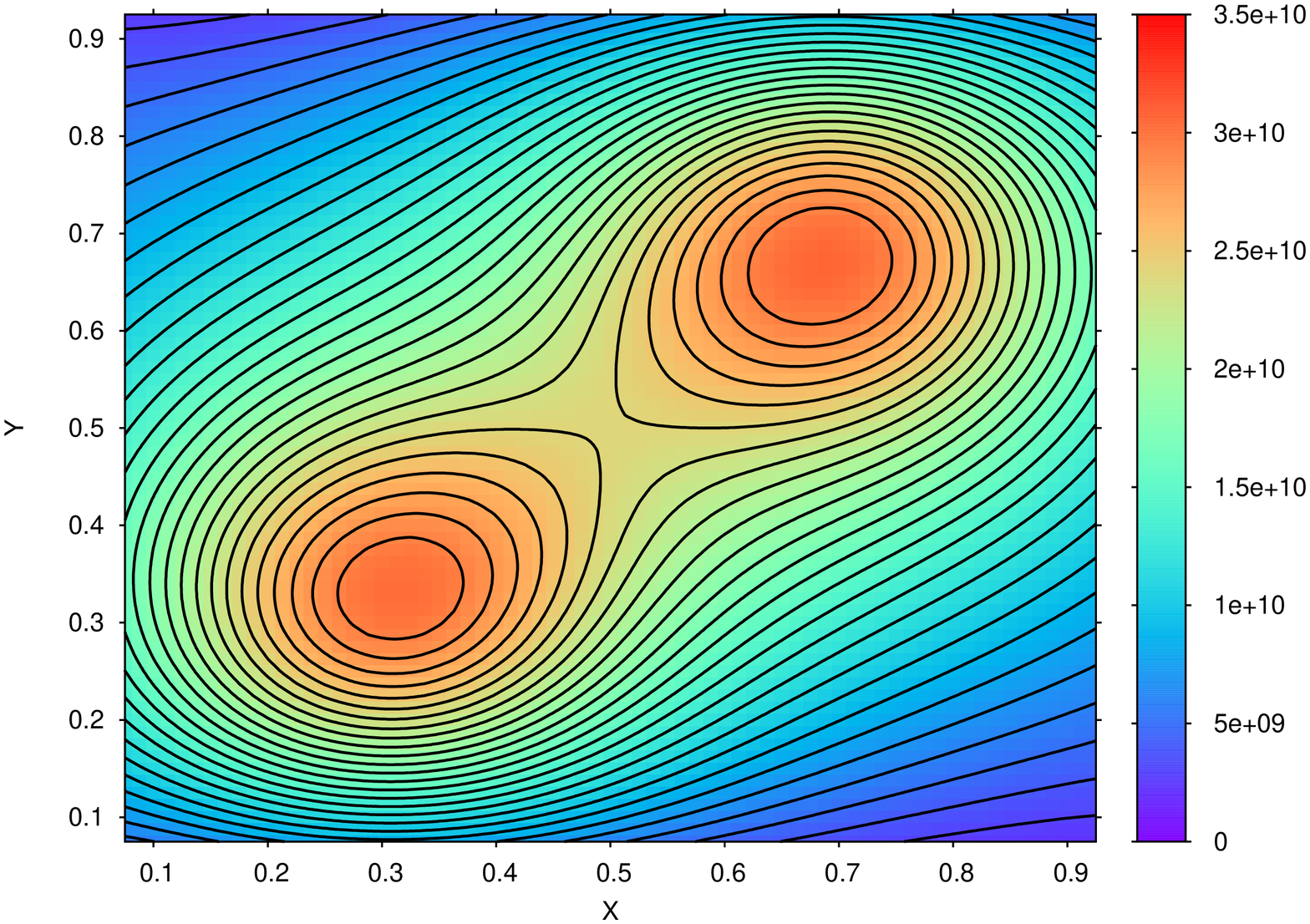}}
        \subfigure[ABC2]{
            \label{fig:2d}
            \includegraphics[width=0.3\textwidth, angle = 270]{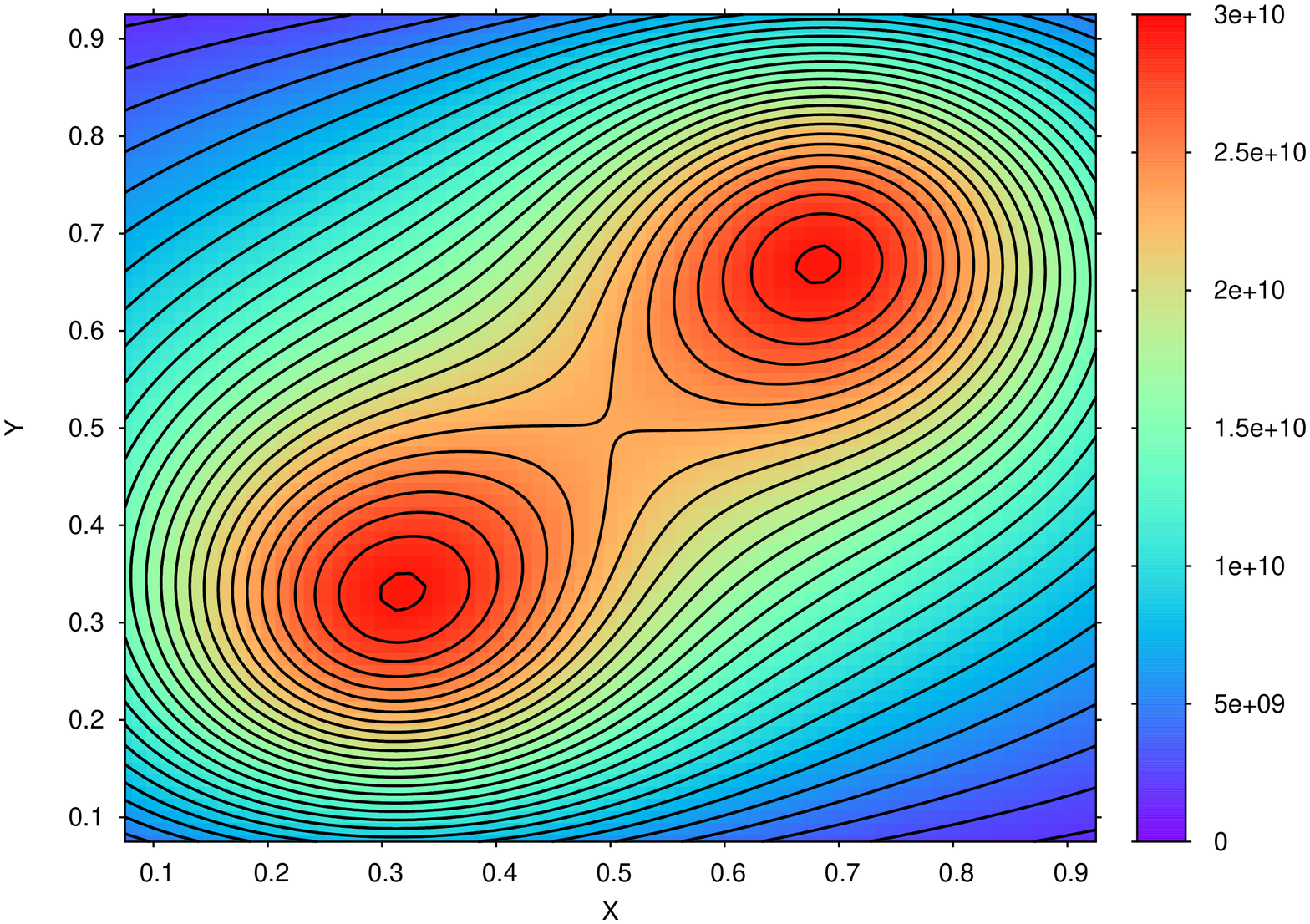}}
    \end{center}
    \caption{
        The potential $V$ for a superposition of two Gaussian charge densities with 
centred at $(0.3,0.3,0.3)$ and $(0.7,0.7,0.7)$. A projection
on the $XY$ plane $z=0.5$ is shown in each case. 
     }
   \label{fig:d2}
\end{figure}

We next consider  a unit cube as before with 
the open faces  at $x = 0$ and $x = 1$m but with a superposition of two
Gaussian charge densities:

\be
\rho(x,y,z) = \begin{cases} \frac{1}{(\sqrt{2\pi})^3\sigma_x\sigma_y\sigma_z} \sum_{i=0}^1 e^{-\frac{(x-x_i)^2}{2\sigma_x^2} -
\frac{(y-y_i)^2}{2\sigma_y^2} - \frac{(z-z_i)^2}{2\sigma_z^2}}  \\
~ & ~  \\
 0 ~~~~~~~~~~~~~~ \mbox{outside the pipe}.  \end{cases} \label{eq:gauss_dipole}
\ee

\noi
with $\sigma_x = \sigma_y = \sigma_z = 1/10$. The results are shown in Fig.~(\ref{fig:d2}).

A comparison of the average relative
errors can again be found in Table 1. 
There is a marked improvement in the performance of ABC2 while Method-1
is consistent.

\subsubsection{A single Gaussian charge density (Case-3)}
\label{subsec:1gaussian}

\begin{figure}[htb!]
     \begin{center}

        \subfigure[Exact]{
            \label{fig:3a}
            \includegraphics[width=0.3\textwidth, angle=270]{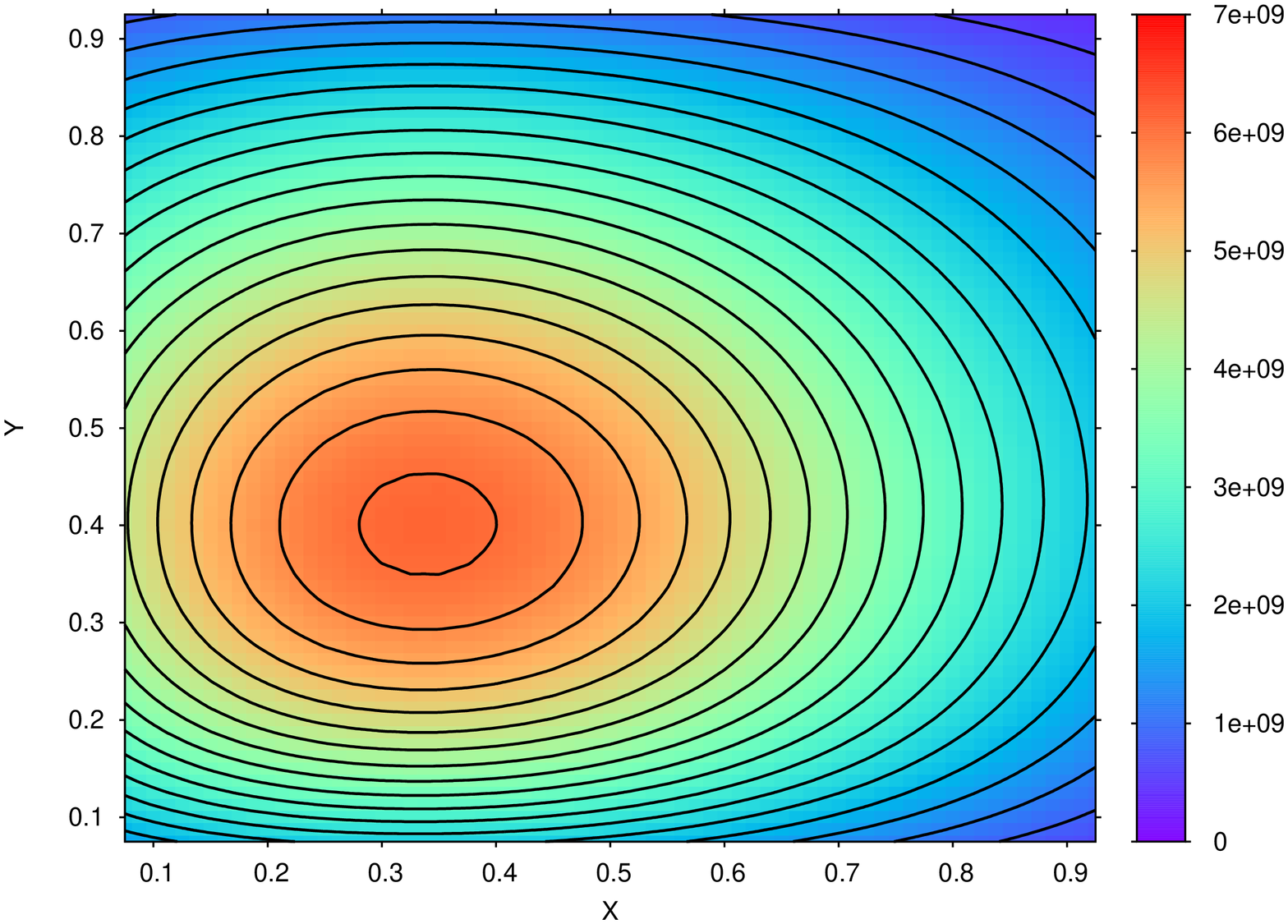}}
        \subfigure[Method-1]{
           \label{fig:3b}
           \includegraphics[width=0.3\textwidth, angle = 270]{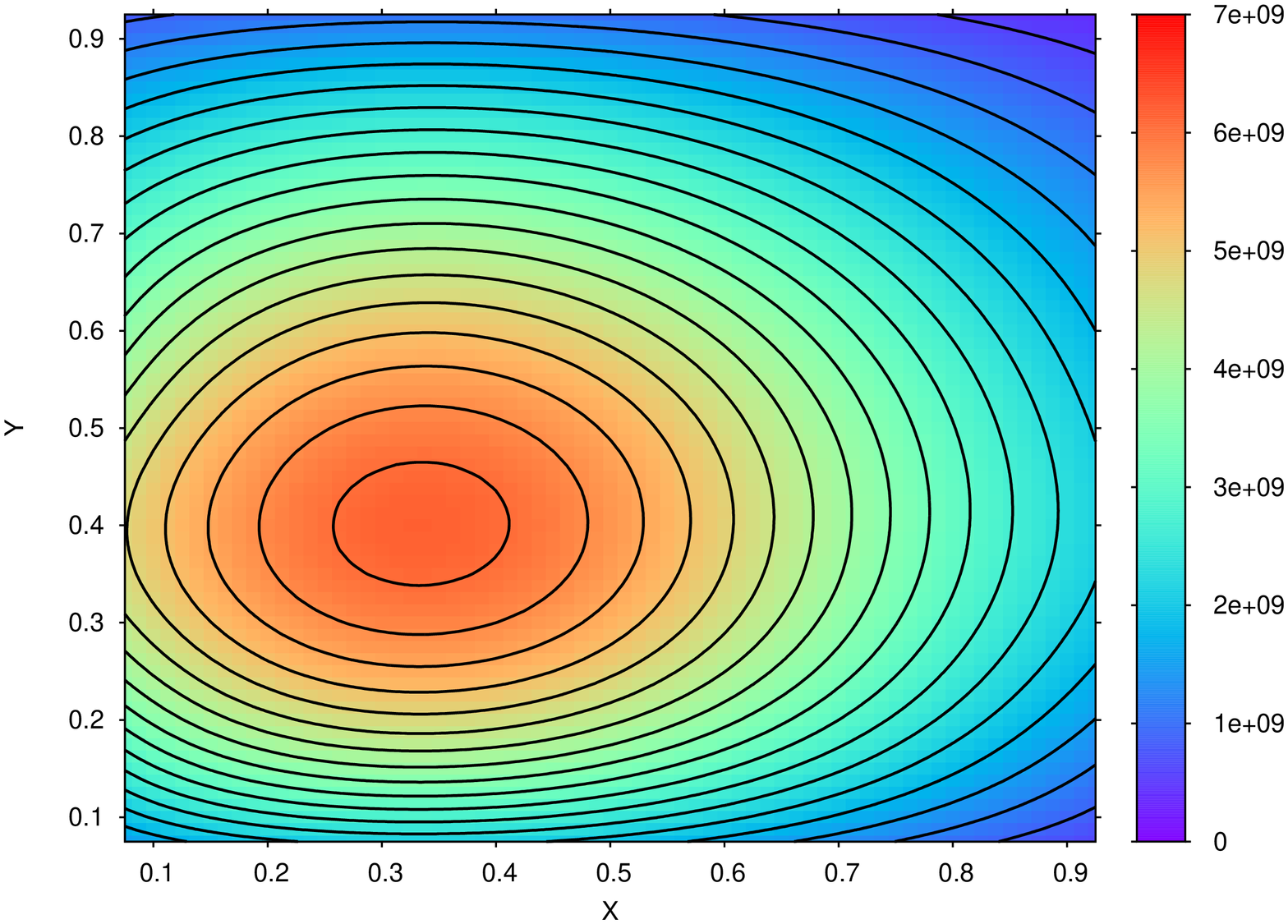}}
        \\ 
        \subfigure[BPM]{
            \label{fig:3c}
            \includegraphics[width=0.3\textwidth, angle= 270]{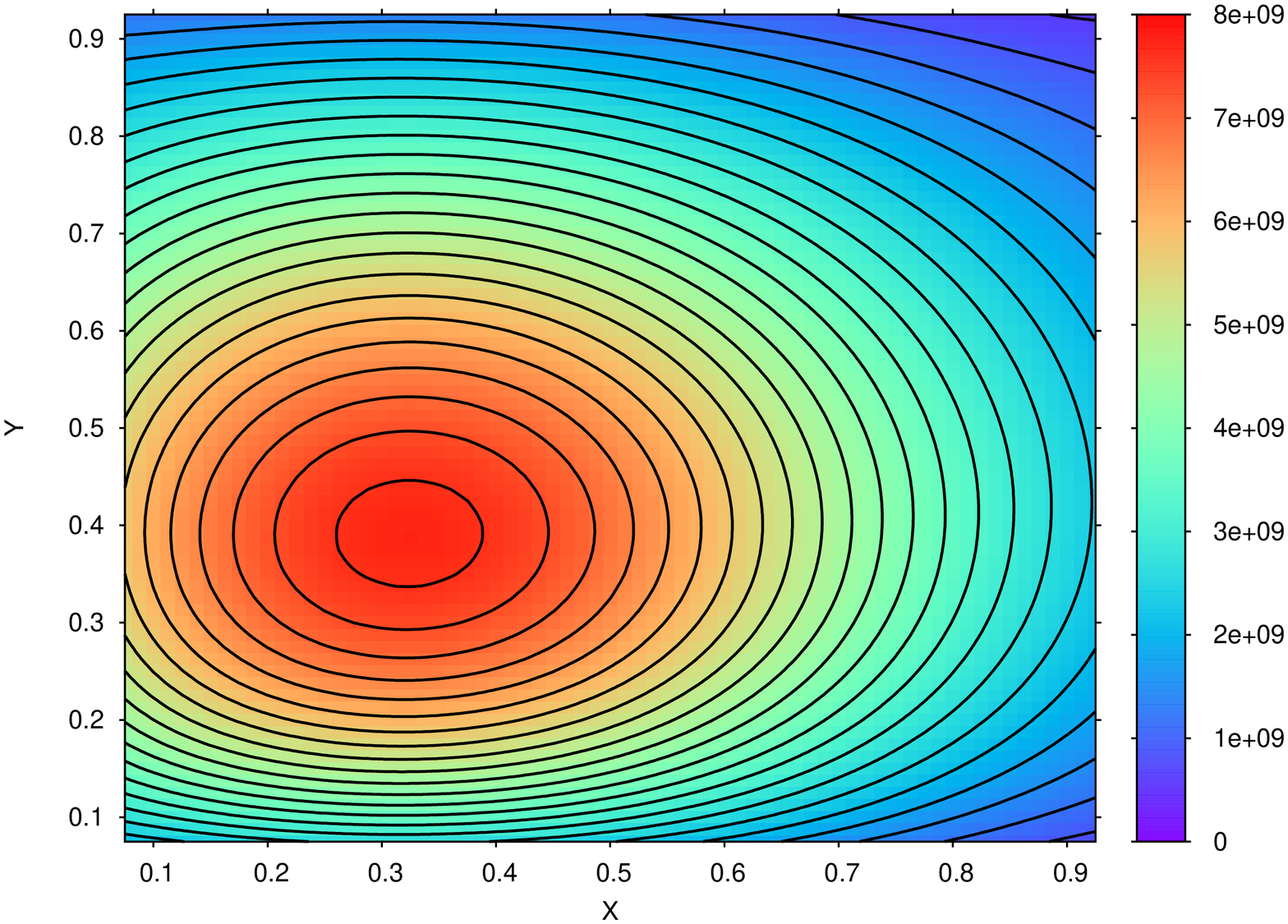}}
        \subfigure[ABC2]{
            \label{fig:3d}
            \includegraphics[width=0.3\textwidth, angle = 270]{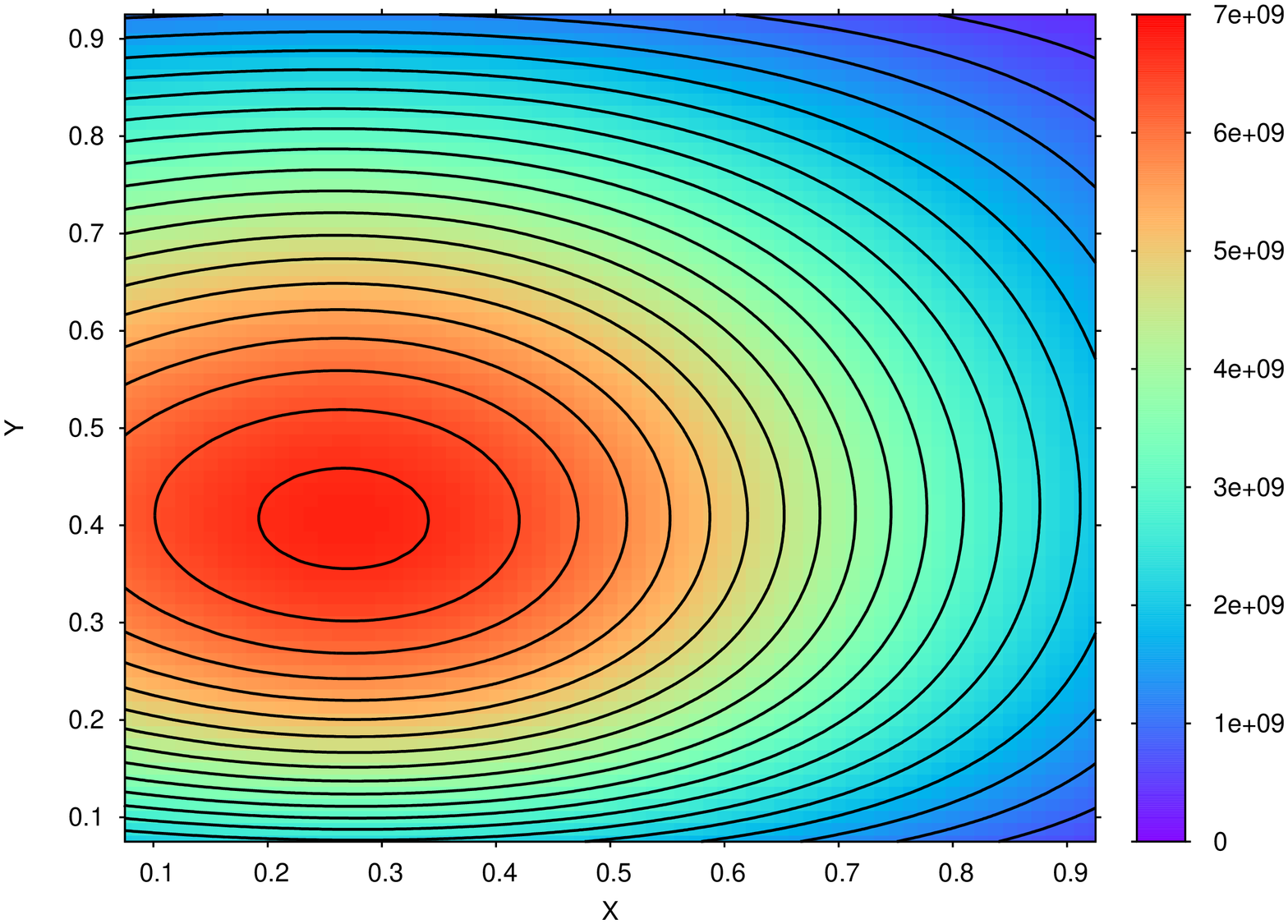}}
    \end{center}
    \caption{
        The potential $V$ for a single Gaussian charge density centred at 
$x_0 = (0.3,0.3,0.3)$. A projection
on the $XY$ plane $z=0.5$ is shown in each case.
     }
   \label{fig:d3}
\end{figure} 

To understand the reason behind the improvement, we consider a single Gaussian 
density placed at (0.3,0.3,0.3) but now 
having $\sigma_x = \sigma_y = \sigma_z =  1/3$.
The performance of ABC2 is no longer as good and the differences can be seen
in Fig.~(\ref{fig:d3}). The boundary potential method (BPM) does not perform well either while
Method-1 remains consistent and fares reasonably well. Table~1 provides
the average relative error in each case.
Clearly, Method-1 performs consistently for all density variations
when the aspect ratio of the
open face is unity while ABC2 performs poorly except in Case-2 where the 
charge is localized.

\begin{figure}[htb!]
     \begin{center}
            \includegraphics[width=0.5\textwidth, angle=270]{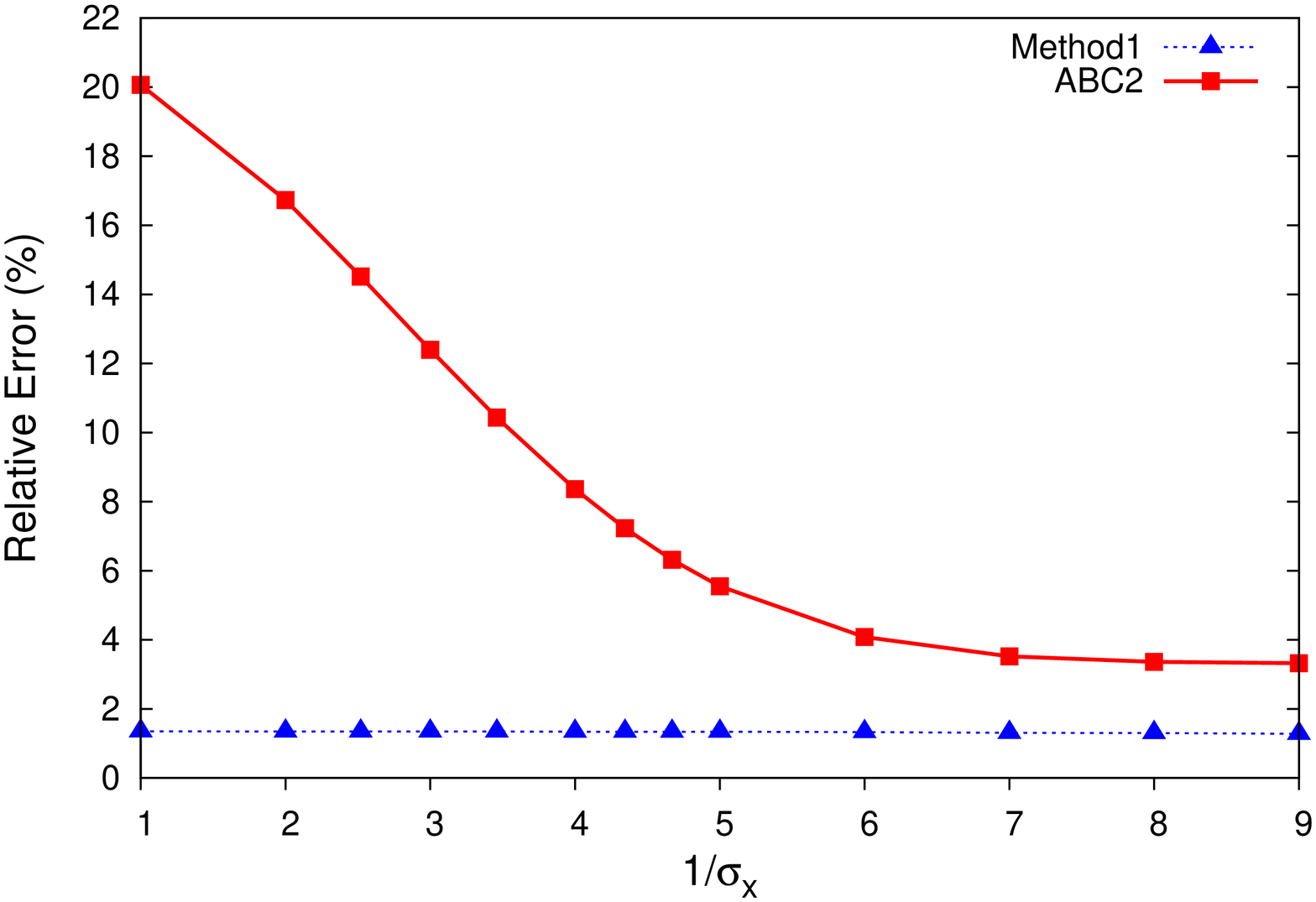}
\end{center}

\caption{The relative error for $L_x = L_y = L_z =1$ with a Gaussian charge density
centred at (0.5,0.5,0.5). The width of the Gaussian varies from $\sigma_x = 1$ till
$\sigma_x = 1/9$. The error reduces for ABC2 as $\sigma_x$ is decreased. For Method-1,
the error increases slightly but remains below that of ABC2.}
\label{fig:difsigma}
\end{figure}

To understand this aspect of ABC2, we consider a single Gaussian
placed in the centre of the open rectangular pipe and vary
the standard deviation $\sigma_x$.
Since the charge density is now at the centre, the effect on both open
faces is now equal. The relative error for ABC2 and Method-1 is
shown in Fig.~\ref{fig:difsigma}. Clearly, the relative error for
ABC2 reduces sharply as the charge is localized and saturates for
small $\sigma_x$ while for Method-1, localization does not change
the relative error substantially.

\begin{figure}[htb!]
     \begin{center}
            \includegraphics[width=0.5\textwidth, angle=270]{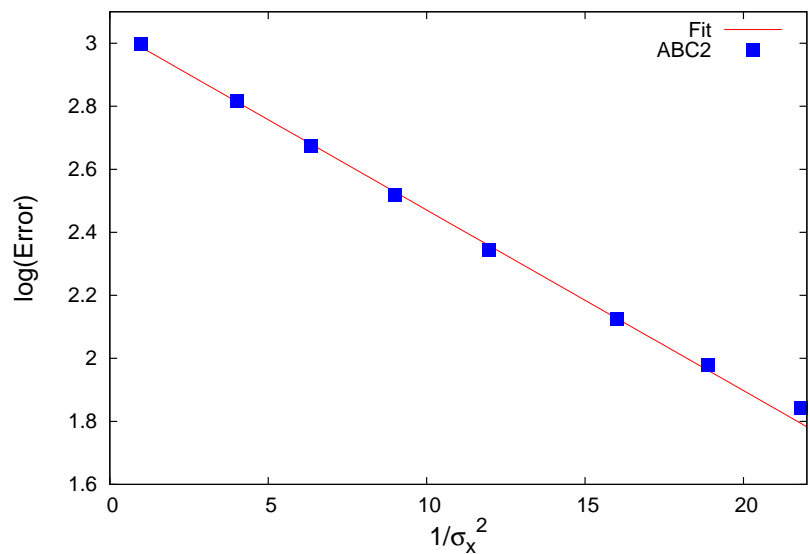}
\end{center}

\caption{The straight line fit shows that the relative error varies as $\exp(-A/\sigma_x^2)$
for large $\sigma_x$}
\label{fig:fitsigma}
\end{figure}

The above observation for ABC2 suggest a relationship between the relative error
and the charge density near the open face relative to the peak density in the 
direction perpendicular to the open face \cite{noeffect}. 
When the length $L_x$ is fixed, we hypothesize that 
the error variation with $\sigma_x$ 
(see Fig.~\ref{fig:difsigma}) depends on how 
the density varies with $\sigma_x$; i.e. 
Error $\sim \frac{1}{\sigma_x}\exp(-A/\sigma_x^2)$ at least for large $\sigma_x$.
To test this, a $\log$(Error) vs $1/\sigma_x^2$ plot is shown in Fig.~\ref{fig:fitsigma}
along with the best fitting straight line. For values of $\sigma_x$ in the
range [1:1/5], the fit is good suggesting that the relative error depends
on the density near the open face relative to the peak density in the
direction perpendicular to the open face.

\begin{figure}[htb!]
     \begin{center}
            \includegraphics[width=0.5\textwidth, angle=270]{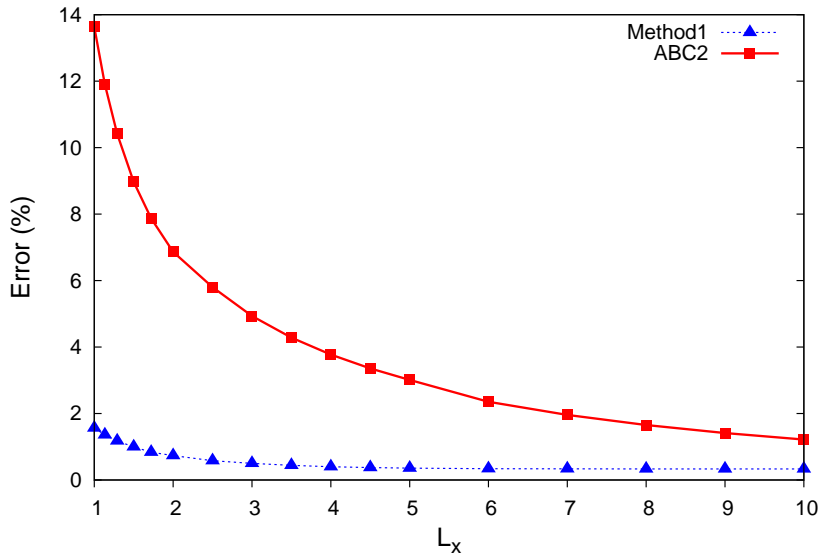}
\end{center}

\caption{The relative error for $L_y = L_z =1$ and the length $L_x$ varying between 1 and 10
for Method-1 and ABC2. The error for ABC2 reduces with length before saturating.}
\label{fig:length}
\end{figure}

\begin{figure}[htb!]
     \begin{center}
            \includegraphics[width=0.5\textwidth, angle=270]{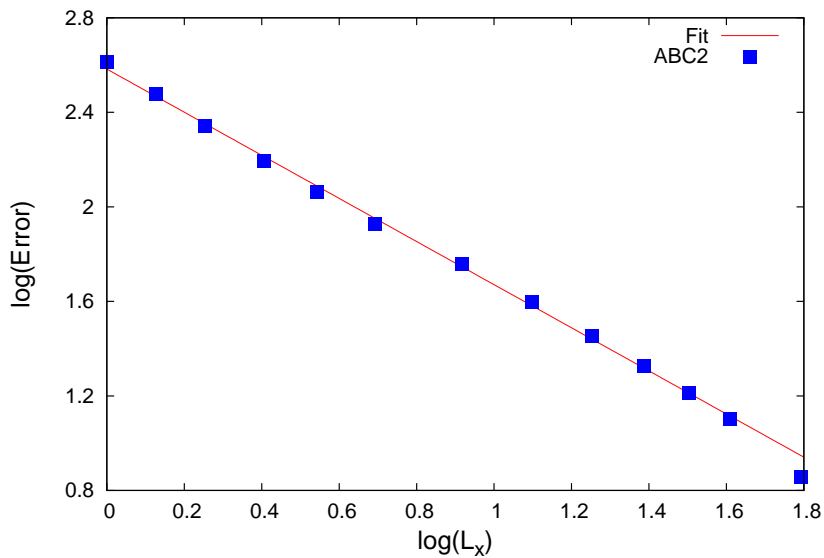}
\end{center}

\caption{The fitted straight line has a slope -1.07 suggesting that the 
the relative error varies as $1/L_x$ for $L_x < 4$.}
\label{fig:fitlength}
\end{figure}

In the above case, while charges were localized on decreasing $\sigma_x$,
the distance from the open face(s) remained invariant.
In order to study the effect of the distance of charge centre from
the open face, we nullify the effect of the dominant 
$\exp(-(x -x_0)^2/2\sigma_x^2)$ term in the Gaussian  
by scaling the point $x_0$ and $\sigma_x$ with $L_x$.
Thus, the relative position of the charge density remains invariant as
the length $L_x$ is increased. In particular, we choose a single 
Gaussian charge density with $\sigma_x = L_x/3$ centred at ($0.3L_x,0.3L_y,0.3L_z$)
and vary $L_x$ from 1 to 10. The error decreases for both ABC2 and Method-1
as shown in Fig.~\ref{fig:length}. This can be ascribed to the 
increase in the distance of the charge centre from the open faces
as $L_x$ is increased.
A plot of $\log(L_x)$ vs $\log$(Error) shows (see Fig.~\ref{fig:fitlength}) 
that the relative error for large $L_x$ decreases inversely as the 
distance from the charge centre to the open face.

The decrease in relative error with increase in distance of the charge centre from the
open face is true for other charge densities as well (including Case-1 where 
the ABC2 error falls from around 20\% at $L_x = 1$ to 1.6\%  for $L_x = 10$).
Note that for both small $\sigma_x$ and large lengths ($L_x$), the error
is dominated by other considerations and saturates.

\subsection{Large Aspect Ratio}

The discussion so far has centred around open faces with unit
aspect ratio such as a circular or square aperture. We shall now
study the suitability of Method-1 and ABC as the aspect ratio of the 
open face is altered keeping the length of the pipe unaltered.
The discretization can now be done in two ways: 
(i) the cell aspect ratio can be unity ($h_x = h_y = h_z$) (ii) the
cell aspect ratio is the same as that of the computational domain 
($h_x:h_y:h_z = L_x:L_y:L_z$). Our studies show that the relative
errors are higher when the cell aspect ratio is unity. For the calculations
presented below, the cell aspect ratio is same as that of 
the computational domain.

\begin{figure}[htb!]
     \begin{center}
            \includegraphics[width=0.5\textwidth, angle=270]{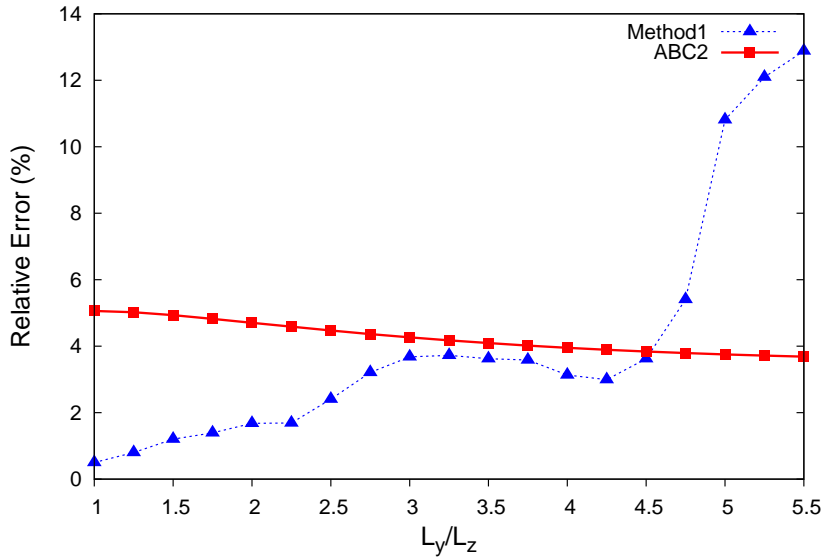}
\end{center}

\caption{The relative error for aspect ratio $L_y/L_z > 1$ and 
$L_x = 3$ for a single Gaussian charge density with $\sigma = L_x/3$. The relative
error grows rapidly for Method-1 beyond $L_y/L_z = 4$.}  
\label{fig:aspect}
\end{figure}

As the aspect ratio is increased (or decreased) from unity, the
domain of convergence of Method-1 decreases and the relative error
increases. A comparison of the change in relative error with 
aspect ratio for ABC-2 and Method-1 is shown in Fig.~\ref{fig:aspect}
for a rectangular tube of length $L_x = 3$ with a single
Gaussian charge density placed at ($0.3L_x,0.3L_y,0.3L_z$)
and having $\sigma_x = L_x/3$, $\sigma_y = L_y/3$ and $\sigma_z = L_z/3$. 
The aspect ratio, $L_y/L_z$ is varied 
such that the area of the open face $L_yL_z = 1$. This ensures that the 
charge density at the open face remains the same as the
aspect ratio is varied since $\sigma_x \sigma_y \sigma_z$ is conserved.
The relative error
using Method-1 in this case rises rapidly for $L_y/L_z > 5$. Thus, Method-1
is suitable in a limited range of aspect ratios.

The domain of convergence of Method-1 generally shrinks rapidly 
for aspect ratios beyond $4$. The Asymptotic Boundary Conditions however
continue to have a large domain of convergence and can be used
for larger aspect ratios. For charges well inside
the computational domain and away from the open boundaries, ABC2
performs consistently well irrespective of the aspect ratio.
We shall therefore focus on higher order and mixed Asymptotic Boundary 
Conditions when the relative density of charges is high close 
to the open boundary. 

To this end, we consider six Gaussian charge densities placed 
such that four of them are at $\sigma_x = L_x/10$ distance from the
open faces while the other two are well inside. With $L_yL_z=1$,
we study the performance of ABC2, ABC3 and a mixture of ABC2 and
ABC3 with 5\% contribution from ABC2. While, ABC3 is
much better than ABC2, the mixture is perhaps the best performer
over the range of aspect ratios considered.

\begin{figure}[htb!]
     \begin{center}
            \includegraphics[width=0.5\textwidth, angle=270]{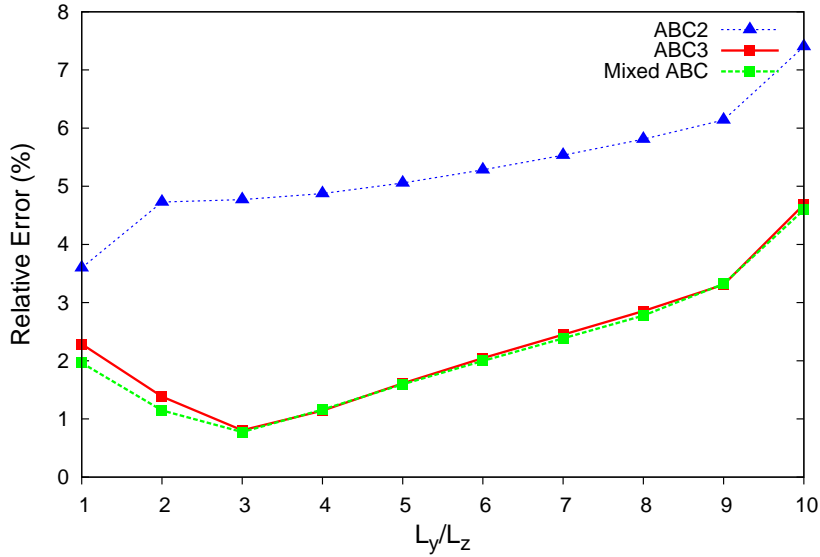}
\end{center}

\caption{The relative error for $L_y/L_z \geq 1$ and $L_x = 3$ for
multiple Gaussian charge density with $\sigma = 1/10$, four of which
are placed near the open faces. }  
\label{fig:aspect1-10}
\end{figure}

For longer lengths however, the significant advantage of ABC3 
decreases and ABC2 performs reasonably well at all aspect ratios
considered.

\section{Discussion and Summary}
\label{sec:summary}

We have considered three methods for solving the Poisson equation 
for open metallic enclosures containing various charge densities. 
Two of these, the ABCs and BPM, are existing methods that
can be directly applied when the computational domain is 
truncated at the open boundaries. We have, in addition, proposed
a non-local truncation (Method-1) based on the solution of the Laplace
equation in the charge-free region outside the metallic enclosure.

It is clear from the numerical results that Method-1, as 
implemented in this paper, is best suited when the aspect ratio of the 
open face is near unity irrespective of the length of the
metallic enclosure. Compared to ABCs, it works especially well
when charges are not localized or when the distance from the
charge centre to the open face is small.
The method uses a 25 term expansion (i.e. 
truncation of the series at $l_{max} = 4$) and for a 
$81 \times 81\times 81$ grid, less than $0.5\%$ of 
the points on the open boundary need
to be matched to determine the unknown expansion coefficients.
It is thus non-local but fast and is consistent in performance with
errors that may be acceptable in many applications especially 
when the interior points are of interest. 
It does however require care in implementation, with the choice of origin 
and the points used for matching, as major factors especially 
in asymmetric geometries.

Our studies also reveal that for ABC2, the charge density localization
in the direction perpendicular to the open face has a direct
bearing on the relative error.
The method is best suited when the
density falls sharply near the open face from its peak
value in the direction perpendicular to the open face. We 
believe this is due the local nature of the boundary condition. 
The relative error also depends sensitively on the 
distance of the charge centre from the open boundary 
especially when the charge density is not localized. Thus for
a density constant in the direction perpendicular to the open
face, the error falls as $1/L_x$ with the length $L_x$.

For larger aspect ratios however, the relative error of Method-1 
rises fast and the convergence domain shrinks rapidly. 
When the aspect ratio of the open face is outside the
range [1/4,4], Method-1 is found unsuitable while 
the local Asymptotic Boundary Conditions (ABC) are stable and
give better results. When charges present are closer to the open faces, 
a combination of 5\% ABC2 and 95\% ABC3 performs consistently 
and may be preferred.




\begin{thebibliography}{00}
\bibitem{noeffect} Increasing the peak density  while retaining the 
shape (e.g. by multiplying the Gaussian by a constant) does
not have any effect on the {\it relative} error.
\end{thebibliography}
\end{document}